\documentclass[11pt,a4paper]{article}

\def\be{\begin{equation}}
\def\ee{\end{equation}}
\def\ba{\begin{eqnarray}}
\def\ea{\end{eqnarray}}
\def\>{\rangle}
\def\<{\langle}
\def\n{\nonumber}
\def\lb{\left[}
\def\rb{\right]}
\def\<{\langle}
\def\>{\rangle}
\def\sc{\scriptsize}
\begin{document}

\title{Comment on the quantum nature of angular momentum
       using a coupled-boson representation}

\author{ILki Kim\hspace{.5mm}\cite{AUTH} and\, Gerald J. Iafrate\vspace*{1.0ex}\\
Department of Electrical and Computer Engineering\\
North Carolina State University,
Raleigh, NC 27695-8617, U.S.A.\\
phone) ++1-919-357-6286; fax) ++1-919-513-1247\\
e-mail) ikim4@ncsu.edu\vspace*{1.0ex}}
%
%
\maketitle
\begin{center}
    {\bf Abstract}
\end{center}
A simple approach for understanding the quantum
nature of angular momentum and its reduction to the classical
limit is presented based on Schwinger's coupled-boson
representation. This approach leads to a straightforward
explanation of why the square of the angular momentum in quantum
mechanics is given by $j(j+1)$ instead of just $j^2$, where $j$ is
the angular momentum quantum number.\vspace*{2.0ex}\vspace*{2.5ex}\\
%
%

Angular momentum in quantum mechanics plays important roles in the
treatment of central force motion, molecular motion, spin
dynamics, and the quantum dynamics of coupled multi-level quantum
systems. Angular momentum oscillators have been a prime basis for
representing atomic species in quantum electronics and quantum
optics. Yet, the ``intuitive'' understanding of the
quantum-mechanical angular momentum and its reduction to the
classical limit is still not often discussed in introductory
textbooks on quantum mechanics. In this paper, we will provide a
simple approach for understanding the quantum nature of angular
momentum and its reduction to the classical limit by considering
the coupled-boson representation of angular momentum, noted by
S{\sc{CHWINGER}} in 1960s \cite{SCH01}.

It is well-known that the square of angular momentum has
eigenvalue equations,
\begin{equation}\label{eq:angular_mom_square}
    \hat{J}^2 \left|jm_j\right\> = j(j+1)\hbar^2
    \left|jm_j\right\>
\end{equation}
and
\begin{equation}\label{eq:angular_mom_z}
    \hat{J}_z \left|jm_j\right\> = m_j \hbar \left|jm_j\right\>\,;
\end{equation}
with $m_j = -j, -j+1, \cdots, j$\,; here $\left|jm_j\right\>$ is
the common set of eigenstates of $\hat{J}^2$ and $\hat{J}_z$ since
$[\hat{J}^2, \hat{J}_z] = 0$. In introductory textbooks,
$\<\hat{J}^2\> = j(j+1)$, instead of $j^2$, is noted as a genuine
quantum-mechanical result without any easy way of understanding
why. In \cite{MIL90}, M{\sc{ILONNI}} provided a very simple and
fancy ``derivation'' of why $j(j+1)$ is the quantum result; it
followed from the symmetry $\<\hat{J}^2\> = 3\,\<\hat{J}_z^2\>$
and assuming the allowance of the only $2j+1$ values of
$\hat{J}_z$ with a well-known sum rule,
\begin{equation}\label{eq:sum_rule}
    \sum_{m_j=-j}^{j} m_j^2\; =\; \frac{1}{3}\, j(j+1)(2j+1)\,.
\end{equation}
But he could not show, as he stated, why the $2j+1$ values only
are allowed.

S{\sc{CHWINGER}} noted that the quantum-mechanical angular
momentum can be obtained by using creation and annihilation
operators of two 1-dimensional harmonic oscillators in the form of
$\hat{J}_{\mu} \propto \hat{a}_1^{\dagger} \otimes \hat{a}_2$,
where $\hat{a}_1$, $\hat{a}_2$ are the annihilation operators of
two linear harmonic oscillators, respectively. This is explicitly
given as follows: \cite{SCH01}
\begin{eqnarray}\label{eq:bosonic_rep}
    \hat{J}_x = \frac{\hbar}{2}
    \left(\hat{a}_1^{\dagger} \otimes \hat{a}_2 +
    \hat{a}_1 \otimes \hat{a}_2^{\dagger}\right)&,&
    \hat{J}_z = \frac{\hbar}{2} \left(\hat{n}_1 - \hat{n}_2\right)\n\\
    \hat{J}_y = \frac{\hbar}{2i}
    \left(\hat{a}_1^{\dagger} \otimes \hat{a}_2 -
    \hat{a}_1 \otimes \hat{a}_2^{\dagger}\right)&,&
    \hat{J} = \frac{\hbar}{2} \left(\hat{n}_1 + \hat{n}_2\right)\,,
\end{eqnarray}
where $\hat{n}_k = \hat{a}_k^{\dagger}\,\hat{a}_k$ with $k=1,2$ is
an occupation number operator of each 1-dimensional oscillator
$k$. By using $\lb \hat{a}_k\,, \hat{a}_l^{\dagger} \rb =
\delta_{kl}$ the operators $\hat{J}_{\mu}$ satisfy the usual
angular-momentum commutation relations. The total occupation
number, $n = n_1+n_2$ with $n_k = 0,1,2 \cdots$ as well as the
operator $\hat{J}$ in (\ref{eq:bosonic_rep}) is a constant of
motion for H{\sc{AMILTON}}ian having no interactions with external
fields (energy conservation).

From this we clearly see a remarkable correspondence between two
linear oscillators and an angular-momentum oscillator. Such a
correspondence was utilized early-on by H{\sc{OLSTEIN}} and
P{\sc{RIMAKOFF}} in connection with the theory of spin waves and
magnon dynamics in ferromagnetic systems \cite{HOL40}; this
correspondence was also well established in the theory of $SU$
symmetries \cite{GRE94}. However, S{\sc{CHWINGER}}'s discourse on
this correspondence shows the seamless general connection between
angular momentum and two independent harmonic oscillators in a
simple and direct fashion \cite{ENG01}. Therefore, the
S{\sc{CHWINGER}} approach is highlighted in this comment for
pedagogical value.

In considering the commutation relation
$\left[\hat{a}_k,\hat{a}^{\dagger}_l\right] =
\epsilon\,\delta_{kl}$ with $\epsilon = 1,0$, respectively, it is
observed that the quantum as well as the classical behavior are
depicted. By using (\ref{eq:bosonic_rep}) with this commutation
relation, we find, after a minor calculation, that
\begin{equation}\label{eq:semi_classical_angular_mom_rel}
    \hat{J}^2\; =\; \hat{J}_x^2\, +\, \hat{J}_y^2\, +\, \hat{J}_z^2\;
    =\; \hat{J}\left(\hat{J}\, +\, \epsilon\,\hat{1}\right)\,.
\end{equation}
Here, we clearly see that the square of the quantum-mechanical
angular momentum must be given by $j(j+1)$ when $\epsilon=1$, with
$j^2$ expressing the classical angular momentum when $\epsilon=0$;
here $j$ is one-half of the total oscillator occupation number,
$j=\frac{1}{2}(n_1+n_2)$. From this property of $j$, it also turns
out that the quantum-mechanical angular momentum is evidently
specified by one of the only allowed values
$j=0,\frac{1}{2},1,\frac{3}{2},2, \cdots$ for all possible values
of $n_1, n_2 = 0, 1, 2, \cdots$. Thus, the allowed quantum numbers
of the angular momentum are integer (e.g. for orbital angular
momenta) or half-integer values as naturally concluded from the
coupled-boson representation. Furthermore, since for $\epsilon=0$,
the operators $\hat{J}_{\mu}$ commute with each other, and the
operators $\hat{n}_j$ in $\hat{J}_z$ and $\hat{J}$ lose their
meaning as number operators, it is naturally expected that for the
classical case $j$ assumes continuous values.

In the quantum case, for a fixed value of $n$, and therefore
$j=\frac{n}{2}$, the $2j+1$ different number states are available
in the sublets of constant $j$\,: $\left(0,2j\right),
\left(1,2j-1\right), \cdots, \left(2j,0\right)$, which can be
characterized by $m_j = -j, -j+1, \cdots, j$ of the $\hat{J}_z$ in
(\ref{eq:bosonic_rep}), respectively. Therefore the total system
with the conservation of the angular momentum $\hat{J}$ in
(\ref{eq:bosonic_rep}) is naturally given by the $(2j+1)$-levels.
This picture clearly provides the answer of why those values of
$m_{j}$ are only allowed for $\hat{J}_z$, which was missing in
\cite{MIL90}. By using the sum rule (\ref{eq:sum_rule}) as in
\cite{MIL90}, we arrive at $\<\hat{J}^2\> = j(j+1) \hbar^2$, thus
obtaining the desired value $j(j+1)$ in a simple manner.

Based on the above result, it is also interesting to consider the
angle between $\hat{J}_z$ and $\hat{J}$ given by
\begin{equation}\label{eq:cos_angle}
    \cos \vartheta_{m_j}\, \equiv\, \frac{J_z}{|\hat{J}|}\,=\,
    \frac{m_j\,\hbar}{\sqrt{j\,(j+\epsilon)\,\hbar^2}}\, =\,
    \frac{m_j/j}{\sqrt{1 + \epsilon/j}}\;,
\end{equation}
independent of $\hbar$. As a maximum and a minimum value of $\cos
\vartheta_{m_j}$ we have $\cos
\vartheta_{j}=\frac{1}{\sqrt{1+\epsilon/j}}$\, and $\cos
\vartheta_{-j}=\frac{-1}{\sqrt{1+\epsilon/j}}$\,, respectively. In
the quantum regime, $\cos \vartheta_{\pm j}$ can never reach the
classically achievable extrema, $\vartheta_{j}=0$,
$\vartheta_{-j}=\pi$, because $\epsilon/j \neq 0$ in observance of
the uncertainty relation; in the classical limit, $\epsilon/j=0$,
and the classical extrema are then realizable. Since $\cos
\vartheta_{\pm j}$ is a universal function of $\epsilon/j$,
independent of $\hbar$, the classical limit can be achieved either
for $\epsilon=0$, $j=\mbox{finite}$ or for $\epsilon=1$, $j \to
\infty$.

In this comment we have simply shown, on the basis of the
coupled-boson representation of angular momentum, why only $2j+1$
values of $\hat{J}_z$ are available in the quantum-mechanical
angular momentum with $\<\hat{J}^2\> = j(j+1) \hbar^2$; also the
quantum nature of angular momentum and its reduction to the
classical limit was easily demonstrated.

\end{document}